\documentclass[12pt,thmsa]{article}
\usepackage{amssymb}

\usepackage{sw20aip}

\input{tcilatex}
\begin{document}

\title{Can quantum vacuum fluctuations be considered real?}
\author{Emilio Santos \and Departamento de F\'{i}sica. Universidad de Cantabria.
Santander. Spain}
\maketitle

\begin{abstract}
The main argument against the assumption that quantum fluctuations of the
electromagnetic field are real is that they do not activate photon
detectors. In order to met this objection I study several models of photon
counter compatible with the reality of the fluctuations. The models predict
a nonlinear dependence of the counting rate with the light intensity and the
existence of a nonthermal dark rate, results which might explain the
difficulty for performing loophole-free optical tests of Bell\'{}s
inequality.

\end{abstract}

\section{\protect\smallskip The vacuum fluctuations of the electromagnetic
field}

The existence of vacuum fluctuations is a straightforward consequence of
field quantization \cite{milonni}. In addition, quantum vacuum fluctuations
have consequences which have been tested empirically. For instance, the
vacuum fluctuations of the electromagnetic field (or zeropoint field, ZPF)
give rise to the main part of the Lamb shift \cite{lamb} and to the Casimir
effect \cite{cm}. The ZPF was proposed in 1912 by Planck, who wrote the
radiation spectrum in the form 
\begin{equation}
\rho \left( \omega ,T\right) =\frac{\omega ^{2}}{\pi ^{2}c^{3}}\left[ \frac{%
\hbar \omega }{\exp \left( \hbar \omega /k_{B}T\right) -1}+\frac{1}{2}\hbar
\omega \right] ,  \label{planck}
\end{equation}
where the second term represents the ZPF. That the thermal spectrum contains
an $\omega ^{3}$ term has been proved by experiments measuring current
fluctuations in circuits with inductance at low temperature \cite{khc}. Of
course, the ZPF term is ultraviolet divergent so that some cutoff should be
assumed, likely at about the Compton wavelength where the fluctuations of
charged Fermi fields (the Dirac electron-positron sea) become important. We
may conclude that the ZPF is well established by both, theoretical and
empirical arguments.

A difficulty for understanding the nature of the ZPF is that we cannot
interpret it na\"{i}vely as a \textit{real }random electromagnetic field, a
kind of universal noise. Indeed it seems to possess properties different
from those to be expected for a classical noise. The two most obvious
differences are that the ZPF apparently does not produce gravitational
effects and that it does not activate photodetectors. With respect to the
first problem, it has been speculated that the quantum vacuum fluctuations
might be at the origin of the cosmological constant, whose nonzero value has
been recently supported by astronomical observations. We shall not be
concerned with gavitational effects in this paper but deal with the second
problem, namely to explain why the ZPF does not activate photodetectors even
in the absence of signals. A common solution to the problem is to say that
the ZPF is not \textit{real}, but \textit{virtual}. But just replacing a
word, real, by another one, virtual, with a less clear meaning is not a good
solution. In the present article I shall attempt to show that the behaviour
of photodetectors can be explained without renouncing to the reality of the
ZPF.

If we compute the intensity of the ZPF, by integrating over frequencies the
second term of eq.$\left( \ref{planck}\right) ,$ we get an extremely large
value. In fact the intensity is of the order of kilowatts per square
centimeter, considering only the visible part of the spectrum. The existence
of such a high intensiy is not the problem, because we might assume that
photodetectors subtract that intensity so that they are sensitive only to
the part which is above ''the sea'' of ZPF. Indeed, the standard quantum
treatment of detection involves the use of normal ordering (of creation and
annihilation photon operators), which may be seen as a formal procedure to
subtract the ZPF. However that formal method cannot be interpreted as
physical, because noise cannot be eliminated by just subtracting the mean.
In order to get a physical interpretation of the subtraction, if we assume
that the ZPF is real, there are two possible ways which might converge. The
first method would be to refine the standard quantum treatment of detection,
the second one is to propose specific detection models resting upon some
plausible assumptions. The second method will be the main subject of the
present paper, but I sketch the first method in the following paragraph.

The standard quantum theory of photodetection starts with an appropriate
light-matter interaction hamiltonian. Hence the probability of photon
absorption is derived using time-dependent perturbation theory and taking,
at some stage, the limit of infinite time. Afterwards it is argued that
microscopic times are so short that the resulting rule ''detection
probability per unit time proportional to the light intensity'' may be used
whatever is the time dependence of the intensity. The procedure usually
provides very good approximation, but it is obviously inconsistent. Indeed,
if the probability per unit time is finite and the time goes to infinity,
the total probability would surpass unity. The problem is that the use of
infinite time intervals hides all the difficulties derived from the ZPF. In
fact, it is trivial to get detection models able to subtract efficiently the
fluctuations of the ZPF if the detection time is large enough, as we shall
comment in the following section. But for actual quantum-optical experiments
(e.g. tests of Bell\'{}s inequality) the detection time is short, typically
of the order of nanoseconds. I conjecture that a refined quantum theory of
detection, which might require short enough detection times and the use of
high order perturbation theory, would give a better understanding of the
quantum vacuum fluctuations.

The point which I want to emphasize is that the difficulties for reaching an
intuitive picture of how detectors subtract the ZPF probably do not derive
from quantum theory itself, but from the use of idealizations like
first-order perturbation theory or infinite detection time. I have
conjectured elsewhere that excesive idealizations might be at the origin of
the difficulties for undersanding intuitively the paradoxical aspects of
quantum physics \cite{es}. Indeed, although simplifications are extremely
useful for calculations, they tend to obscure the physics. In the case of
photon counting we might attempt to understand the removal of the ZPF by
working a refined quantum theory of detection, but here I shall use the
alternative route of studying detection models without explicit appeal to
quantum theory. However I stress that these models do not necessarily
contradict quantum theory.

\section{Critical analysis of recent detection models}

Several models of photodetection have been proposed recently resting upon
the idea that there exists a ''detection time'', T, independent of the light
intensity and such that the probability of a count depends on the radiation
(including the zeropoint field, ZPF) which enters the detector during the
time T \cite{kaled}, \cite{crs}. It proved necessary to ''filter'' the
incoming field by means of some temporal and spatial Fourier transform in
order to reduce the noise. A general feature of these models is that the
''filtered'' intensity of the ZPF has a gaussian distribution, whose
dispersion I shall label $\sigma .$ Finding a suitable filtering is the
difficult problem in the construction of a model because it would not work
if $\sigma $ is not smaller than the typical light intensities to be
detected. In the particular model of \cite{crs} the quantity $\sigma $ was
rather sensitive to the assumed depth of the active zone of the
photodetector, a fact which has been proved incompatible with experiments 
\cite{genovese}. However, there are experiments which refute, not only that
particular model, but the whole family of models involving a fixed detection
time T (see above) as we show in the following.

In the mentioned family of models we start calculating the probability
distribution of the total radiation energy which enters the detector during
the time T. (As the energy is proportional to the beam intensity, I shall
not distinguish between energy and intensity in the present section.) In the
most simple case, where the incoming beam consists of a signal with constant
intensity, I$_{s},$ superimposed to the ZPF, the probality distribution is 
\begin{equation}
\rho (I)=K\exp \left( \frac{(I-I_{0}-I_{s})^{2}}{2\sigma ^{2}}\right) ,
\label{ro}
\end{equation}
K being a normalization constant and $I_{s}$ the mean intensity of the ZPF.

In order that the model predictions do not depart too much from the quantum
predictions, the detection probability should be roughly proportional to the
intensity (except at high intensities where there should appear saturation
effects). Then it is plausible to assume that the probability, Q, of a count
when the energy I has entered the detector during the ''detection time'' T,
is given by 
\begin{equation}
Q(I)=\xi (I-I_{0})\Theta \left( I-I_{0}-I_{m}\right) ,  \label{p}
\end{equation}
I$_{m}$ ( I$_{m}$ \TEXTsymbol{>} 0) being a threshold, $\Theta ()$ the
Heavside step function and $\xi $ a constant related to the detector
efficiency. The first factor, if alone, would give the quantum rule
''counting probability proportional to the intensity'', but the second
factor is necessary in order to ensure that Q $\geq 0,$ as it should Q being
a probability.

After that the counting rate as a function of the incoming deterministic
part of the intensity, I$_{s}$, is 
\begin{equation}
R\left( I_{s}\right) =\frac{1}{T}\int \rho (I)Q(I)dI.  \label{rate}
\end{equation}
Whatever are the values of the parameters I$_{m}$ and $\sigma $ (the value
of I$_{0}$ is irrelevant) the curve $R(I_{s})$ has the following features.
At high intensity we have $R$ $\propto $ $I_{s}$ in agreement with the
quantum rule, but for $I_{s}=0$ there is some unavoidable ''dark rate''. We
may reduce the dark rate by increasing the threshold I$_{m}$, but if I$_{m}$
is high the departure of $R(I_{s})$ from linearity is big. This is because
the curve lies well below the linear asymptote (tangent at infinite)
whenever I$_{s}$ is not substantially larger than $\sigma ,$ that is 
\begin{equation}
R\ll \xi I_{s}\text{ for }I_{s}\lesssim \sigma ,\text{ where }\xi
=\lim_{I_{s}\rightarrow \infty }\left( \frac{R}{I_{s}}\right) .
\label{lowsignal}
\end{equation}

When one goes from singles counts to (correlated) coincidence counts, it can
be shown that the following inequality holds \cite{cs} (R$_{1}$ labels
singles rate and R$_{12}$ coincidence rate) 
\begin{equation}
R_{12}<\left( 1+\left( \frac{\sigma }{I_{s}}\right) ^{2}\right) TR_{1}^{2}.
\label{inequality}
\end{equation}
This implies that either T should be very large (i.e. T $\gg
R_{12}/R_{1}^{2} $ ) or the correlation in coincidence counts may be
observed only when the intensity is so low that the function $R_{1}(I_{s})$
is nonlinear (i.e. $I_{s}$ $\ll \sigma $, see eq.$\left( \ref{lowsignal}%
\right) $). High correlation exists, for instance, when there is high
visibility in the polarization correlation measured in order to test the
Bell inequalities. This conclusion is not compatible with empirical
evidence. In fact Fig.4 of the recent article by Kurtsiefer et al. \cite{kow}
shows an essentially linear counting rate, R, as a function of the
intensity, I$_{s}$, along two orders of magnitude, and the paper reports a
visibility of 98\% for the highest intensity of this linear region. In order
that these results are compatible with $\left( \ref{inequality}\right) $ we
should have T $\gtrsim $ 1/$R_{1}$, which is absurd.

We conclude that the whole class of models considered in the present section
are incompatible with empirical evidence.

\section{Detection models with varying detection time}

Instead of fixing the detection time, T, we may assume that a count is
produced when the radiation energy accumulated in the detector surpases some
threshold. This means that when the photocounter is ready to detect (this
will happen some ''dead time'' after a count is produced, but we will
neglect the dead time here), the detector begins to accumulate the radiation
energy entering in it. If I$_{tot}$(t) is the total intensity entering the
detector at time t, the accumulated energy at time T will be 
\begin{equation}
E(T)=A\int_{0}^{T}I_{tot}(t)dt,  \label{energy}
\end{equation}
where A is the entrance area of the detector (in the following we shall put
A = 1 for the sake of simplicity). Our assumption is that a new count will
be produced when E(T) surpasses some threshold E$_{t}$ (this threshold may
depend on the value of T). According to our essential assumption that the
ZPF has the same nature as the signal and it is indistinguishable from it,
we may write the intensity in the form

\begin{equation}
I_{tot}(t)=I_{0}+I(t),  \label{inot}
\end{equation}
where I$_{0}$ is the mean of the ZPF entering the detector. Taking eqs.$%
\left( \ref{energy}\right) $ and $\left( \ref{inot}\right) $ into account we
define the threshold in the form 
\begin{equation}
E_{t}=I_{0}T+E_{m}.  \label{thres}
\end{equation}

Thus our model assumption is: \textit{a detection event is produced at a
time T, after the previous count, when T is such that } 
\begin{equation}
\int_{0}^{T}I(t)dt=E_{m},  \label{model}
\end{equation}
\textit{where I(t) is the radiation intensity entering the detector, once
the average intensity of the ZPF has been subtracted, and E}$_{\mathit{m}}$%
\textit{\ is a parameter characteristic of the detector.}

The use of eq.$\left( \ref{model}\right) $ is cumbersome due to the
fluctuations of the ZPF (and maybe also fluctuations of the signal). Indeed
constructing a detailed detection model on the basis of that equation would
require using the theory of ''first passage time'' for the stochastic
process $I(t)$. In the following we shall solve, using rough approximations,
the case of a deterministic signal with constant intensity I$_{s}$. In this
case I(t) is the sum of the ZPF part, I$_{ZPF}$(t), plus I$_{s}$ and eq.$%
\left( \ref{model}\right) $ may be written

\begin{equation}
E_{m}=E_{0}\left( T\right) +I_{s}T\simeq \Sigma \sqrt{T}+I_{s}T,E_{0}\left(
T\right) \equiv \int_{0}^{T}I_{ZPF}(t)dt.  \label{mod}
\end{equation}
There are two approximations in the second equality. The first one is to
treat the stochastic process $I_{ZPF}(t)$ as a Wiener (''white noise'') one.
The approximation may be justified if the coherence time of the incoming
radiation is much shorter than T, which we shall assume here. The second
approximation is to assume that the first time at which the stochastic
process $E_{0}(T)$ takes some fixed value, say $E_{1},$ is proportional to
the time needed for the equality 
\[
\left\langle E_{0}{}^{2}\right\rangle =E_{1}^{2}
\]
to hold. These approximations give a first passage time proportional to $%
\sqrt{T}$, which lead us to eq.$\left( \ref{mod}\right) $giving a detection
model which contains two parameters, $E_{m}$ and $\Sigma $, characteristic
of the detector. From eq.$\left( \ref{mod}\right) $ it is easy to get T,
whence the counting rate becomes 
\begin{equation}
R=\frac{1}{T}=\frac{4I_{s}^{2}}{\left( \sqrt{\Sigma ^{2}+4I_{s}E_{m}}-\Sigma
\right) ^{2}}.  \label{rat}
\end{equation}

The interesting case is that of high intensity, which suggests an expansion
in powers of the small parameter  $\Sigma /\sqrt{I_{s}E_{m}}$. We get
\begin{equation}
R\simeq \frac{I_{s}}{E_{m}}+\frac{\Sigma \sqrt{I_{s}}}{\sqrt{E_{m}}^{3}}+%
\frac{\Sigma ^{2}}{2E_{m}^{2}}.  \label{r}
\end{equation}
From this result we see that there exists a counting rate even without
signal, given by the last term, which we may considered as a part of the
dark rate. The existence of a fundamental dark rate is an unavoidable
consequence or the fluctuation of the ZPF. On the other hand the function $%
R(I_{s})$ is almost linear with a slowly decreasing slope. This behaviour
fits qualitatively the results reported in the experiment (see Fig.4 of the
quoted paper \cite{kow}. )

If our model is correct we are able to make some specific predictions which
might be tested experimentally. For instance a fit of our curve eq.$\left( 
\ref{r}\right) $ to the empirical data \cite{kow} gives our prediction for
the dark rate to be 
\[
R_{dark}\gtrsim \frac{\sigma ^{2}}{2E_{m}^{2}}\approx 30s^{-1},
\]
for that experiment. We cannot make a comparison with the empirical dark
rate because the is not given in the paper \cite{kow}. Another
straightforward prediction of our model is that the coincidence rate should
decrease more quickly than the single rate when the intensity decreases.
This is what happens in Fig. 4 of the mentioned paper \cite{kow}. Indeed
from the highest to the lowest intensity the slope of the singles count
curve, $R(I_{s}),$ decreases by about 25\%, but the decrese in the slope of
the coincidence curve is almost 40\%. The explanation of this behaviour is
that the ''detection time'' T increases as the signal intensity decreases
(see eq.$\left( \ref{mod}\right) )$. Now, the fluctuations of the energy
accumulated during a time T are less relevant as the time T increases. On
the other hand the correlation between two beams is just a correlation of
the fluctuations\cite{cms} (this is true at least for the correlated beams
produced in the process of parametric down conversion,\ PDC, which was used
in the commented experiment .) Consequently it is expected that the measured
correlation is weaker as the correlated beams are less intense, although we
are not able to give a quantitative prediction for the moment. More accurate
predictions will be possible when a detailed detection model is made along
the lines presented in this section, a work which is in progress.

I finish this section stressing that our model goes beyond, although not
necessarily against, quantum mechanics. In fact, the analysis of
quantum-optical experiments is typically made starting from the quantum
predictions for the ideal case. Afterwards these predictions are modified by
introducing some empirical parameters like efficiencies of detectors,
polarizers, etc., in order to fit actual laboratory situations. The
nonidealities are attributed to technical imperfections of the set-up and
are not analyzed further. Here I have studied a part of the nonidealities,
namely those relative to detectors. This is why I say that our model goes
beyond standard quantum mechanics.

\section{Conclusion}

Our analysis shows that quantum vacuum fluctuations of the electromagnetic
field (or ZPF) are a possible source of nonidealities in the behaviour of
optical photon counters. This is specially important when it is necessary to
measure coincidence counting rates with short time windows, as is frequent
in quantum optical experiments (e.g. optical tests of Bell\'{}s inequality).
In contrast they are probably irrelevant for measurements lasting for long
times, as is usually the case in astronomical observations.

The main effect of the ZPF on photon counters is that the response to the
incoming intensity becomes nonlinear, as is shown by our eq.$\left( \ref{r}%
\right) $. In particular the ZPF gives rise to a fundamental, nonthermal,
dark rate and a decrease in the effective efficiency of the detector with
increasing beam intensity. I may conjecture that the nonidealities will
dramatically increase when the detection efficiency is high. In fact, if our
eq.$\left( \ref{r}\right) $ is correct, at least as a rough approximation,
then the increase in efficiency could be reached only by lowering the
parameter $E_{m}$, which will lead to a big departure from the ideal
behaviour. It must be taken into account that the parameter $\Sigma ,$ which
essentially measures the intensity of the ZPF fluctuations, could not be
reduced too much with the design of the photocounters.

Our results provide a possible explanation for the difficulties of
performing loophole-free tests of Bell\'{}s inequality using optical
photons. As is well known all performed experiments suffer from the
''detection loophole'' \cite{laloe} and I conjecture that the cause might be
the existence of fundamental nonidealities in the behaviour of photon
counters \cite{crs}.

\textbf{Acknowledgement. }I acknowledge financial support from DGICYT,
Project No. PB-98-0191 (Spain).


\begin{thebibliography}{99}
\bibitem{lamb}  Prof. Lamb has expressed the opinion that his experiment
provided an empirical proof that the vacuum is not empty. This is related to
his statement that the photon is the quantum of the electromagnetic field,
but not a particle, see W. E. Lamb Jr., \textit{Appl. Phys.} \textbf{B 60},
77 (1995). The point is important for the present paper because it stresses
the difference between photon counting and detection of true particles, like
electrons or atoms. Indeed it is not difficult to detect atoms with
practically 100\% efficiency and no noise, but this is not the case with
photons.
\end{thebibliography}
\end{document}